\documentclass[12pt,preprint]{aastex}

\newcommand\beq{\begin{equation}}
\newcommand\eeq{\end{equation}}

\begin{document}
 
\title{Young Crab-like pulsars and luminous X-ray sources
in starbursts and optically dull galaxies}

\author{Rosalba Perna\altaffilmark{1,}\altaffilmark{2} and 
Luigi Stella\altaffilmark{3}}

\affil{1. Department of Astrophysical Sciences, Princeton
University, Princeton, NJ, 08544}
\affil{3. INAF - Osservatorio Astronomico di Roma, 
Via Frascati 33, I-00040 Rome, Italy}

\altaffiltext{2}{Spitzer fellow}

\begin{abstract}

Recent {\em Chandra} observations of nearby galaxies have revealed a
number of ultraluminous X-ray sources (ULXs) with super-Eddington
luminosities, away from the central regions of non-active galaxies.
The nature of these sources is still debated.  We argue that a
fraction of them could be young, Crab-like pulsars, the X-ray
luminosity of which is powered by rotation. We use the pulsar birth parameters
estimated from radio pulsar data to compute the steady-state pulsar
X-ray luminosity distribution as a function of the star formation rate
(SFR) in the galaxy. We find that $\sim 10\%$ of optically dull
galaxies are expected to have a source with $L_x \ga 10^{39}$ erg/s,
while starbursts galaxies should each have several of these sources.
We estimate that the X-ray luminosity of a few percents of galaxies is
dominated by a single bright pulsar with $L_x\ga 10^{39}$ erg/s,
roughly independently of its SFR.  We discuss observational
diagnostics that can help distinguish the young pulsar population in
ULXs.

\end{abstract}


\section{Introduction}

It has been known since the ROSAT (and Einstein) surveys that the
X-ray luminosity function of normal galaxies extends up to 
luminosities $\sim 10^{42}$ erg/s (e.g. Fabbiano 1989
for a review).  This high-luminosity tail
 was soon attributed to
galaxies characterised by conspicuous regions of intense star
formation activity (``starbursts").  Young (and short-lived) X-ray
binary systems, mainly binaries with a high mass donor star (``high
mass X-ray binaries", HMXBs), are expected to be tracers of star
formation. Therefore the higher X-ray luminosity of starburst galaxies
compared to normal galaxies should result from their larger population
of young X-ray binaries. Our Galaxy contains some $\sim 80$ high mass
X-ray binaries with a persistent luminosity $\ga 10^{35}$ ergs/s;
their integrated luminosity is in the $10^{38}$ ergs/s range.  The
bulk of this luminosity derives from the higher luminosity sources,
which in our galaxy do not exceed $\sim 10^{38}$ erg/s. On the other
hand, the integrated X-ray luminosity of starburst galaxies is often
in the $\sim 10^{40}-10^{41}$ erg/s range. The sensitivity and spatial
resolution of the ROSAT and Einstein telescopes proved insufficient to
resolve the X-ray emission of starbursts into its main constituents.
Yet images of nearby galaxies suggested in several cases the
presence of highly super-Eddington stellar-mass sources ($\sim 10^{40}$ erg s$^{-1}$
level).

The superior spatial resolution and sensitivity afforded by the 
X-ray telescope on board {\it Chandra} has recently yielded two important 
new pieces of information. Firstly, the images of a some nearby starbursts
(e.g. those in NGC4579 (Eracleous et al. 2002), NGC3256 (Lira et al. 2002) 
and the Antennae (Zesas et al. 2002)) demonstrated that 
an important fraction (up to a half) of their X-ray luminosity derives 
from a few very bright individual X-ray sources, the luminosity of which 
is $\sim 10^{39.5}-10^{40}$ and $\sim 10^{40}-10^{40.5}$ erg/s in the 
case of the Antennae and NGC 3256, repectively. 
Secondly, optical follow-up studies of sources in deep {\it Chandra} images
revealed a few early type galaxies that show no sign of 
nuclear activity and yet emit X-rays likely from an individual source
at the $\sim 10^{41}-10^{42}$ erg/s level (e.g. Hornschmeier et al. 2003;
Norman et al. 2004; Fabbiano \& White 2003
for a review and Ptak \& Colbert 2004 for a summary of the statistical
properties of galaxies with ULXs). 

A variety of models have been proposed to explain the origin of these
ultraluminous X-ray sources. These include, among others, accreting
intermediate mass black holes (IMBH) possibly in binaries,
microquasars, transient super-Eddington accretors, background BL Lac
objects, young supernova remnants (SNRs) in extremely dense
environments, beamed X-ray binaries (Fabbiano 1989; Marston et
al. 1995; Colbert \& Mushorsky 1999; Makishima et al. 2000; King et
al. 2001; Karet et al. 2001; Begelman 2002; Miller et al. 2003). All the
observational evidence gathered so far does not appear to point
uniquely to any of the above scenarios. ULXs might actually be a
heterogeneous population.

In this paper, we argue that a sizeable fraction of these
ultraluminous X-ray sources are likely to be young, Crab-like pulsars,
the X-ray emission of which is powered by their rotational energy rather than
by accretion. A number of these sources is indeed expected in any
star-forming region galaxy; these sources would not require
any ``special'', unusual condition to output a large X-ray luminosity
when they are very young.  Here we compute their X-ray luminosity distribution,
based on the birth parameters determined from pulsar radio studies;
we then discuss observational diagnostics that can help identify them
among the population of ULXs.

\section{Pulsar X-ray luminosity}

Most isolated neutron stars are X-ray bright throughout all their lifetime:
at early times, their X-ray luminosity is powered by rotation
(e.g. Michel 1991; Becker \& Trumper 1997a and references therein);
after an age of $\sim 10^3 - 10^4$ yr, when the star has 
slowed down sufficiently, the dominant X-ray source becomes the internal energy
of the star, and finally, when their internal heat is exhausted,
accretion by the ISM continues to make them shine again, although to a 
low luminosity level (e.g. Blaes \& Madau 1993; Popov et al. 2000; Perna
et al. 2003).

What is of interest for this work, is the origin of the X-ray luminosity at
early times, when rotation is the main source of energy. 
During a Crab-like phase, relativistic
particles accelerated in the pulsar magnetosphere are fed to a synchrotron 
emitting nebula\footnote{It appears that all energetic pulsars with
$\dot{E}_{\rm rot}\ga 4\times 10^{36}$ erg/s possess a pulsar
wind nebula (Gotthelf 2004).},  the emission of which is  characterized 
by a powerlaw spectrum. Another important contribution is the 
pulsed X-ray luminosity (about 10 \% of the total in the case of the Crab) 
originating directly from the pulsar magnetosphere. 

Observationally there appears to be a correlation between the
rotational energy loss of the star, $\dot{E}_{\rm
rot}=I\Omega\dot{\Omega}$, and its X-ray luminosity, $L_x$. This
correlation was first noticed in a small sample of radio pulsars by
Seward \& Wang (1998), and later examined in more detail by Becker \&
Trumper (1997). They found that a sample of 27 pulsars detected with
ROSAT in the 0.1-2.4 keV band could be well described by the relation
$L_x(\dot{E}_{\rm rot})\approx 10^{-3} \; \dot{E}_{\rm rot}\;$.  This
correlation appears to hold over a wide range of rotational energy
losses, encompassing different emission mechanisms of the pulsar
(see above). In the high $L_x$ regime of interest for this work,
however, the X-ray emission is dominated by rotational energy
losses. A more appropriate energy band for the correlation
$L_x-\dot{E}_{\rm rot}$ is above $\sim 2$ keV, where the contribution 
of atmospheric emission due to the
internal heat of the star (as well as the uncertainty in the spectral
fitting due to the interstellar absorption) is reduced. The
$L_x-\dot{E}_{\rm rot}$ correlation in the 2-10 keV band was first examined by
Saito (1998) for a small sample of pulsars.  The most comprehensive
investigation up to date has been performed by Possenti et
al. (2002). They found, for a sample of 39 pulsars, that the
$L_x-\dot{E}_{\rm rot}$ correlation is best fit by the relation
\beq
L_x = 10^{-15.3}\,(\dot{E}_{\rm rot}/{\rm erg\; s^{-1}})^{1.34}\; {\rm erg\; s^{-1}}.
\label{eq:Lx}
\eeq
Although empirically established,
there is no physical reason for which Eq.(\ref{eq:Lx}) should break down for
values of $\dot{E}_{\rm rot}$ larger than the currently observed range. We therefore
assume that it holds also for larger values of $\dot{E}_{\rm rot}$.  
Given the scatter in  the data, we allow for a scatter in the
relation (\ref{eq:Lx}) by assigning, to a pulsar of rotational energy
$\dot{E}_{\rm rot}$, an X-ray luminosity drawn from a log-Gaussian  distribution of
mean given by Eq(\ref{eq:Lx}) and standard deviation (in log) $\sigma=0.5$. 
Our results are not particularly sensitive to the precise value of $\sigma$. 
It should be noted that the extrapolation of Eq.(\ref{eq:Lx}) at luminosities much higher
than the observed range could lead to values of $L_x>\dot{E}_{\rm rot}$. To
prevent this from happening, we further impose the condition that the efficiency
$\eta_x\equiv L_x/\dot{E}_{\rm rot}$ be always $\le 1$. Note that, in their sample,
Possenti et al. found this efficiency to  be as high as 0.8, and that $\eta_x$
appears to be an increasing function of $\dot{E}_{\rm rot}$. 

The rotational energy loss of the star, under the assumption that it is dominated
by magnetic dipole losses, is given by
\beq
\dot{E}_{\rm rot}\simeq \frac{B^2\sin^2\theta\,\Omega^4\,R^6}{6c^3}
\;, 
\label{eq:Edot}
\eeq
where $R$ is the radius of the neutron star, which we assume
fixed at a value of 10 km,
$B$ its magnetic field, $\Omega=2\pi/P$ its angular velocity, and 
$\theta$ the angle between the magnetic and spin axes. 

In order to compute the X-ray luminosity distribution of a population
of young pulsars, the magnetic fields and the initial periods of the
pulsars need to be known. Modeling the intrinsic properties of the
Galactic population of pulsars, with the purpose of inferring their
intrinsic properties, has been the subject of extensive investigation
(e.g. Gunn \& Ostriker 1970; Phinney \& Blandford 1981; Lyne et
al. 1985; Stollman 1987; Emmering \& Chevalier 1989; Narayan \&
Ostriker 1990; Lorimer et al. 1993; Cordes \& Chernoff 1998). The most
recent, comprehensive analysis, based on large-scale 0.4 GHz pulsar
surveys, has been carried out by Arzoumanian, Cordes \& Chernoff
(2002; ACC in the following).  Here we use their inferred parameters
under the assumption that spin down is only caused by dipole radiation
losses and there is not a significant magnetic field decay.  They find
that the initial magnetic field strength (taken as Gaussian in log)
has a mean $\langle \log B_0[G] \rangle =12.35$ and a standard
deviation of $0.4$, while  the initial birth period
distribution (also taken as a log-Gaussian), is found to have a mean $\langle \log
P_0(s) \rangle =-2.3$ with a standard deviation greater than $0.2$.
Here we adopt the value 0.3 while discussing the dependence of our
results on $\sigma$ (see \S 3.1). Note that ACC take $\sin\theta=1$
in Eq.(\ref{eq:Edot}), and therefore this is what we assume here to be
consistent with their parameter determination.  

The spin evolution of the
pulsars is then simply given by: \beq P(t) = \left[P_0^2 +
\left(\frac{16\pi^2 R^6 B^2}{3Ic^3}\right)t\right]^{1/2}\;,
\label{eq:spin} 
\eeq
where $I$ is the moment of inertia of the star.

ACC derive that the pulsar birth rate $\dot{N}$ consistent
with their initial parameters is of one pulsar every 760 yr.
This value, which corresponds to a star formation rate (SFR) of
$\sim 0.2 M_\odot$ yr$^{-1}$ for a Salpeter IMF, is a factor
of a few smaller than most other estimates in the literature.
Here, we will show our results for this value of the SFR as well as for 
larger values that are appropriate for starburst galaxies.

\section{Bright X-ray sources from a population of young pulsars}

\subsection{Montecarlo simulation of the luminosity function}

The question we address here is the following: given a galaxy where the
pulsar birth rate is $\dot{N}$, what is the number of sources with 
luminosity larger than $L_X$ at any given time?

Firstly, we wish to determine the probability distribution of a pulsar
to have a given luminosity {\em at birth}. We determine this distribution
by performing a Montecarlo simulation where, for any given pulsar, the
initial period and magnetic field are randomly generated from the 
corresponding distributions described in \S2. For that given set of parameters,
Eqs. (\ref{eq:Edot}) and (\ref{eq:Lx}) are then used to compute the rotational
energy loss and the corresponding X-ray luminosity. Figure 1 shows the resulting
probability distribution, i.e. the fraction of pulsars with an X-ray
luminosity at birth larger than $L_X$, for three different values (0.3,0.4,0.5)
of the width $\sigma_{P_0}$ of the initial period distribution (for
which ACC were able to derive only the lower limit $\sigma_{P_0}>0.2$.    
While ACC cut their lower period distribution at 0.1 ms, we cut it
at 0.5 ms, corresponding to a maximum velocity $\Omega_{\rm max}\sim 12500$ rad/s,
consistent with the break up speed (Cook et al. 1994) for typical NS equations
of state (Wiringa et al. 1988). 
The curves in Figure 1 are obtained by randomly
generating 20000 pulsars. As it can be seen from the 
figure, a substantial fraction of the pulsar population has an X-ray luminosity at birth
which is super-Eddington. This fraction clearly increases with the
width $\sigma_{P_0}$ of the spin period distribution.
Here we adopt the value $\sigma_{P_0}=0.3$, which
is consistent with the lower limit derived by ACC while being on the
more conservative side. 

In order to make a comparison with current observations, we need to
answer the question posed at the beginning of the section, that is
finding the number of sources with a given luminosity in any galaxy at
any given time. Besides depending on the initial birth parameters
$P_0$, $B$ (and their time evolution), this number also depends on the
birth rate of pulsars.  As discussed above, we will use here both the
pulsar birth rate derived by ACC, as well as higher rates as expected
in starburst galaxies.  We perform a Montecarlo simulation of the
young pulsar population as follows: for a given birth rate $\dot{N}$,
we take the total simulation time to be $T\gg 1/\dot{N}$, so that
steady state is reached (typically $T\sim 10^4 1/\dot{N}$). The number
of pulsars generated during this time is $N_{\rm puls}=T/(1/\dot{N})$,
with ages randomly drawn from a flat distribution between 0 and $T$.
The birth parameters $P_0$, $B$ are drawn from the distribution by ACC
described in \S2; the periods of the pulsars are evolved, for the pulsar ages,
according to Eq.(\ref{eq:spin}), and the luminosity as a function of
time of each pulsar is estimated according to Eq.(\ref{eq:Lx}).
For every choice of $\dot{N}$, the results are the average over 2000
different montecarlo realizations.

The four curves in Figure 2 show the resulting X-ray luminosity distribution
in a galaxy with a pulsar birth rate of $1/10\,,1/50\,,1/200\,,1/760$
yr$^{-1}$. The luminosity function
is described by a roughly ``universal'' function, i.e. a power law
with index $\alpha \approx -0.4$, whose normalization is proportional
to the pulsar birth rate (or equivalently the SFR). An interesting
comparison is the one between the X-ray luminosity function of the pulsars
and that of the HMXBs. The latter, as shown by Grimm et al. (2003), also has
an almost ``universal' shape, which, in cumulative form, can be
described by the function $N(>L_{38})\approx 5.4$ SFR$(M_\odot {\rm
yr}^{-1})[L_{38}^{-0.61} -210^{-0.61}]$.  The normalization depends on
the SFR and for the Galaxy they quote a value of $\sim 0.25 M_\odot$
yr$^{-1}$ as found from a combination of different SFR indicators.

Figure 3 shows a comparison between the integrated HMXB and pulsar counts 
for a galaxy with a SFR$\sim 1 M_\odot$/yr.  
The relative normalization between the two populations
is calibrated to be the same as that for the Galaxy, i.e. so that a pulsar
rate of 1/760 yr$^{-1}$ corresponds to a SFR rate of 0.25
$M_\odot$/yr.  The HMXB population generally dominates the X-ray
luminosity at sub-Eddington luminosities; however at higher
luminosities, where HMXBs drop out, the luminosity function of pulsars
takes over.  Our results show that, for a pulsar birth rate typical of
our Galaxy, $\sim 7\%$ of galaxies are expected to have at least one
source with luminosity $\ga 10^{39}$ erg/s, and $\sim 0.3\%$ with
luminosity $\ga 10^{40}$ erg/s. Starburst galaxies, with 
SFR $\sim 10-20 M_\odot/$yr, are expected to each have at least one source with
$L_x\ga 10^{40}$ erg/s.  Note that our simulation for 
the ACC pulsar rate (about 1/4 of the
value shown in Fig.3), predicts that there should be $\sim 1$
Crab-like pulsars ($L_X\sim 10^{36}-10^{37}$ erg/s) in our Galaxy,
consistently with observations.

Another question of interest is the fraction of galaxies in which the
{\em total} X-ray luminosity is dominated by that of a single young
pulsar source.  To address this issue, we ran 5000 random realizations
of the steady-state pulsar population in a galaxy with a SFR similar
to that of the Galaxy. We kept track of all the cases where the
luminosity of a single source was $\ge 90\%$ of the total X-ray
luminosity of the galaxy due to all the pulsar sources together.  The
fraction of galaxies whose luminosity is dominated by a that of a
single source according to the criterion above is shown in Figure 4 as
a function of the luminosity of the source.  If all the X-ray sources
in a galaxy were pulsars, then $\sim 10\%$ of galaxies would be
dominated in X-rays by a single bright source. However, for a given
SFR, one expects a number of HMXBs to be present. We therefore
addressed the same question including in the total luminosity of the
galaxy also the contribution from HMXBs.  To this purpose, for each
random realization of the pulsar population in the galaxy, we also
randomly generated $N_{\rm HMXB}$ from their own luminosity
distribution, where $N_{\rm HMXB}\sim 80$ is the number of HMXBs
corresponding to the SFR of the Galaxy (Grimm et al.  2003).  Figure 4
shows that, although the luminosity function of the HMXBs is a few
times higher than that of the young pulsar population at luminosities
$\la 10^{40}$ erg/s, $\sim 2\%$ of galaxies will be dominated in
X-rays by a single young pulsar.  While the number of sources per
galaxy with luminosity above a certain value roughly scales with the
SFR (Fig.3), the number of sources that dominate the total luminosity
of the galaxy is roughly independent of the SFR, given the almost
universal shape of the luminosity function of both pulsars and
HMXBs. In this calculation we did not include the contribution to the
luminosity function of low-mass X-ray binaries (LMXBs), since their
overall contribution in star forming regions\footnote{In regions of
very low SFR the LMXBs might dominate due to their much longer
lifetime (Persic et al. 2004). However, due to their sharp cutoff
at $L_x\sim 10^{39}$ erg/s they do not substantially 
influence the statistics of ULXs.}
is well below that of HMXBs, and that of an hypothetical population of
intermediate-mass black holes, which has not been observationally
determined as yet.

It should also be noted that a young pulsar is probably surrounded by a bright
SNR, and that the X-ray emission from a young SNR can be substantial,
sometimes comparable to or larger than that of the associated pulsar
(Immler \& Lewin 2002). If the luminosity of the SNR is
super-Eddington and larger than that of the pulsar, then the previously
proposed SNR-ULX connection would hold. However, a recent study of young SNR
luminosites by Bregman et al. (2003) showed that it is unlikely for
young SNRs to make up a substantial fraction of the ULX population
(see also Zesas et al. 2002 for spectral studies).  At any event,
separating a possible contribution to the luminosity from the SNR
could only be done spectroscopically, since the angular size of the
X-ray emitting shell is expected to be well in subarcsec range. On the
contrary a thermal-like spectral component could testify to the
presence of the expanding SN shell. 
Note that the SN ejecta are likely to be optically thick
soon after the SN explosion. In the case of SN 1987A, Fransson
\& Chevalier (1987) estimated an optical depth for absorption of $\sim 6$ about
1 yr after the explosion at $E\sim 1$ keV (for an expansion velocity
of the star core star $V_c=2430$ km/s). The opacity scales
as $(V_c t)^{-2}$ and therefore Type Ib/c SNe, which have larger velocities,
would be more optically thin. Moreover, if the material were clumpy
the opacity would also be reduced. To make a conservative estimate of
the reduction in the source counts due to the opacity of the remnant,
we ran our simulations assuming that no radiation from the pulsar
could be detected in its first 10 yr of life. We found that the
number of sources with $L_x>10^{39}$ erg s$^{-1}$
is reduced by $\sim 20\%$ for $\dot{N}=1/10$ yr$^{-1}$,
and $\sim 13\%$ for $\dot{N}=1/760$ yr$^{-1}$ (other cases are in between). 
What is interesting to note is that, if a source is observed at two
different times while the remnant is becoming optically thin, then
an {\em increase} in flux would be observed, which is not otherwise
expected in our model (see also \S3.2).

\subsection{Observational diagnostics of the young pulsar population}

We summarize below some observational diagnostics that can help identifying 
young rotation powered pulsars among the population of ULXs. 

{\em (a) X-ray spectra, variability and polarization}

Young, rotation-powered pulsars display power law energy spectra. In a sample 
of pulsars studied with {\em Chandra}, Gotthelf (2003) found their 
photon indices to be in the range $0.6<\Gamma_{PSR}<2.1$ (while their
associated pulsar wind nebulae had $1.3<\Gamma_{PSR}<2.3$). 
It should be noted that several ULXs have powerlaw
spectral components consistent with this range of values (Wang et al.
2004). Pulsations might be detected with high time resolution, high
throughput, X-ray light curves of ULXs. The super-Eddington range of
the distribution is produced by the subset of sources with smaller
periods and larger magnetic fields. More specifically, in our Montecarlo
simulations of the pulsar population, we found that $\sim 90$\% of the
sources with $L_x\ga10^{39}$ erg s$^{-1}$ has periods in the range $\sim$ 2$-$9 ms,
and magnetic fields $\sim 4\times 10^{11}-8\times 10^{12}$ G (with larger
periods correlating with larger magnetic fields, obviously).   
No appreciable intrinsic variability is expected on timescale of days to
months. Aperiodix flux variability, if observed in these sources,
would likely be due to external causes. The passage of a cloud along
the line of sight to the source might even be expected for a very
young pulsar in the dense enviroment of the SN debris or the
progenitor star wind. Spectral analysis would easily uncover the
signature of variations due to increased photoelectric absorption.
Other spectral changes, due to intrinsic variability of the source are
not expected. A small, constant decline in luminosity is
however predicted due to the pulsar spin down.  For a spin down by
pure dipole, the X-ray luminosity declines with time as
$L_x=L_{x,0}(1+t/t_0)^{-2}$, where $t_0\equiv 3Ic^3P_i^2/B^2R^6(2\pi)^2
\sim 65\, {\rm yr}I_{45}B^{-2}_{12}R_{10}^{-6}P^2_{i,{\rm ms}}$,  
having defined $I_{45}= I/(10^{45}$~g~cm$^2$), $R_{10}= R/(10\; {\rm km})$,
$P_{i,{\rm ms}}=P_i/(1\;{\rm ms})$.  
For $t\la t_0$ the flux does not vary significantly. In our simulations, we 
kept track of the value of $t_0$ and found that, for pulsars with 
$L_x\ga10^{39}$~erg~s$^{-1}$, about a third of them has an age $t>t_0$,
implying that their luminosity is expected to decay with time over a timescale 
of years. 

Finally, based on the analogy with the Crab, a high polarization degree ($\sim
20\%$) is expected (Weisskopf et al. 1978). Prospects for measuring
X-ray polarization in relatively low-flux sources will dramatically
improve in the future (Costa et al. 2001).

{\em (b) Association with star forming regions and young SN
remnants}

Newly born pulsars, even if endowed with large kick velocities, have
not had the time to travel far from their birth sites, and therefore
are expected to be mostly found in star forming regions.  {\em
Chandra} observations of a sample of ULXs in nearby galaxy (Roberts et
al 2004) seems to support the association between these sources and
regions of intense star formation.  On a more general, statistical level,
given that young pulsars directly trace the SFR, we expect that
active, starbursts galaxies will contain a much larger fraction of
ULXs than optically dull galaxies, where there is no significant star
formation going on. For an optically dull galaxy with a SFR $\sim 0.2
M_\odot$/yr, we predict that there is a $\sim 7\%$ probability of
finding a source with $L_x\ga 10^{39}$ erg/s and a $\sim 0.3\%$
probability of finding a source with $L_x\ga 10^{40}$
erg/s. Correspondingly, for an active, starburst galaxy with a SFR
$\sim 10 M_\odot$/yr, we expect to find $\sim$ a few sources with
$L_x\ga 10^{39}$ erg/s and $\sim 1$ source with $L_x\ga 10^{40}$
erg/s.
Finally, the presence of a young SNR ($T_{SNR}\la$ several tens of
yr) surrounding the source is a fundamental ingredient of the young
pulsar scenario.  There is at least one case where the association
between a ULX and a SNR has been established (Roberts et al. 2003).

{\em (c) Binarity}

A large number of stars are in binary sistems.  Whether a binary
survives the first SN explosion will mostly depend on the kick
velocity of the NS and the mass of the companion; for high mass
companions up to several tens of percent of binaries are expected to
survive the first SN explosion (Brandt \& Podsiadlowski 1995; Kalogera
1998). Therefore, a sizeable fraction of young pulsars is expected to
be in a binary system. On the other hand, while binarity is required
in most models of ULXs, for the young pulsar scenario it is not a
requirement as the source of the X-ray luminosity is the rotational
energy of the star.

{\em (d) Optical and radio emission}

Young pulsars are also detected in radio and often in optical
(e.g. Becker \& Trumper 1997 for a review).  For the Crab, the ratio
$L_{\rm opt}/L_x$ is $\sim 10^{-2}$, while for Vela it is $\sim 10^{-3}$.
The statistics of optical detections in young pulsars is too small
to yield predictions about the intensity of the optical radiation 
as a function of the X-ray one. However, if the ratio remains of the
same order as for Crab and Vela, then significant optical emission
is expected, although prospects for detection appear promising
only for sources in relatively close by galaxies.

The greatest majority of young X-ray pulsars are also radio emitters.
Although the radio luminosity generally anti-correlates with age
(Camilo et al. 2002), the radio luminosity for the youngest pulsars is
still many orders of magnitude below the X-ray one ($L_{\rm
radio}/L_x\sim 10^{-5}-10^{-10}$ at 1GHz), and radio detection in
external galaxies is extremely difficult.  The radio flux from a
source associated with a young pulsar would instead be dominated by
the radio emission from its young SN remnant.

{\em (e) {Possible disk component}}

A young pulsar can possibly have a surrounding disk either from
fallback material from the supernova explosion, (e.g. Chatterjee et
al. 2000; Alpar 2001; Menou et al. 2001), or from a companion if it is
in a binary. The emission spectra of fallback disks around young
neutron stars (e.g. Perna, Hernquist \& Narayan 2000)
depends on a combination of parameters, and in particular on the magnetic field
of the star and its spin, and on the accretion rate. The inner radius of 
the disk is generally determined by the magnetospheric radius; however
the brightest young pulsars are the ones with shortest periods, and for those
the disk, if at all, will most likely extend to the light cylinder.
For a millisecond pulsar, the light cylinder is only a few times the 
radius of the star, and the inner part of the disk will emit in X-rays.
It is probably difficult to separate this component from the
X-ray bright pulsar, and therefore what might be potentially detectable
is the longer wavelength radiation produced in the outer parts of the
disk. 

{\em (f) Statistical properties and caveats}

The model described in \S 3 makes specific predictions for the
pulsar luminosity function. In principle, this function could
be compared against the detected X-ray point sources, after all the
other contributions have been identified and substracted out. 
We chose not to perform a detailed comparison with X-ray source
statistics, as our results, in particular for the high $L_x$ tail
of the distribution, rely on the still uncertain details of the
initial period distribution and on the $L_x-\dot{E}_{\rm rot}$
correlation at high values of  $\dot{E}_{\rm rot}$. In both cases,
we have adopted the results from the most updated and comprehensive
studies up to date, but we should still caution about these uncertainties.
Also, there might be corrections if the X-ray emission is somewhat
beamed. Brazier \& Johnston (1999) estimated the beaming fraction 
of the young radio pulsars to be at least 50\%, and noted that
nearly all of the radio detections are also X-ray detection, while
the converse is not true (implying that the X-ray beaming fraction is
probably higher in X than in radio),  
while Possenti et al. (2002) suggested that the
scatter in the $L_x-\dot{E}_{\rm rot}$ correlation itself might indeed
be due to viewing effects. At any event,
given the above, it is interesting to note the converse, which is, that once the nature
of all ULXs has been indentified, then the fraction of these
that are indeed associted with young pulsars can be used as an
independent constraint to calibrate the properties of the young pulsar
population. 
  
\section{Summary}

The nature of ultra luminous X-ray sources is a current issue of
debate. In this paper we have proposed that a fraction of them could
be associated with rotation-powered pulsars.  We have performed a
Montecarlo simulation which uses the pulsar birth parameters inferred
from radio data analysis, together with an emphirical correlation
between X-ray and spindown luminosity, to predict the
X-ray luminosity function of this population. We have found that a few
percent of optically dull galaxies are expected to possess young
pulsars with highly super-Eddington luminosity, while starbursts
galaxies can each have several of them.  We have discussed other pieces of
observational evidence that can help identify these sources among the
observed population of ULXs.

\acknowledgments
LS acknowledges useful discussions with F. Fiore and M. Vietri. 
We thank B. Gaensler and the referee for insightful comments on the
manuscript.
This work was partially 
supported by the Italian Space Agency (ASI) and the Ministero
della Ricerca Scientifica e Tecnologica (MURST). RP ackowledges
support from a Lyman Spitzer, Jr. fellowship at Princeton university.


\clearpage

\begin{figure}[t]
\plotone{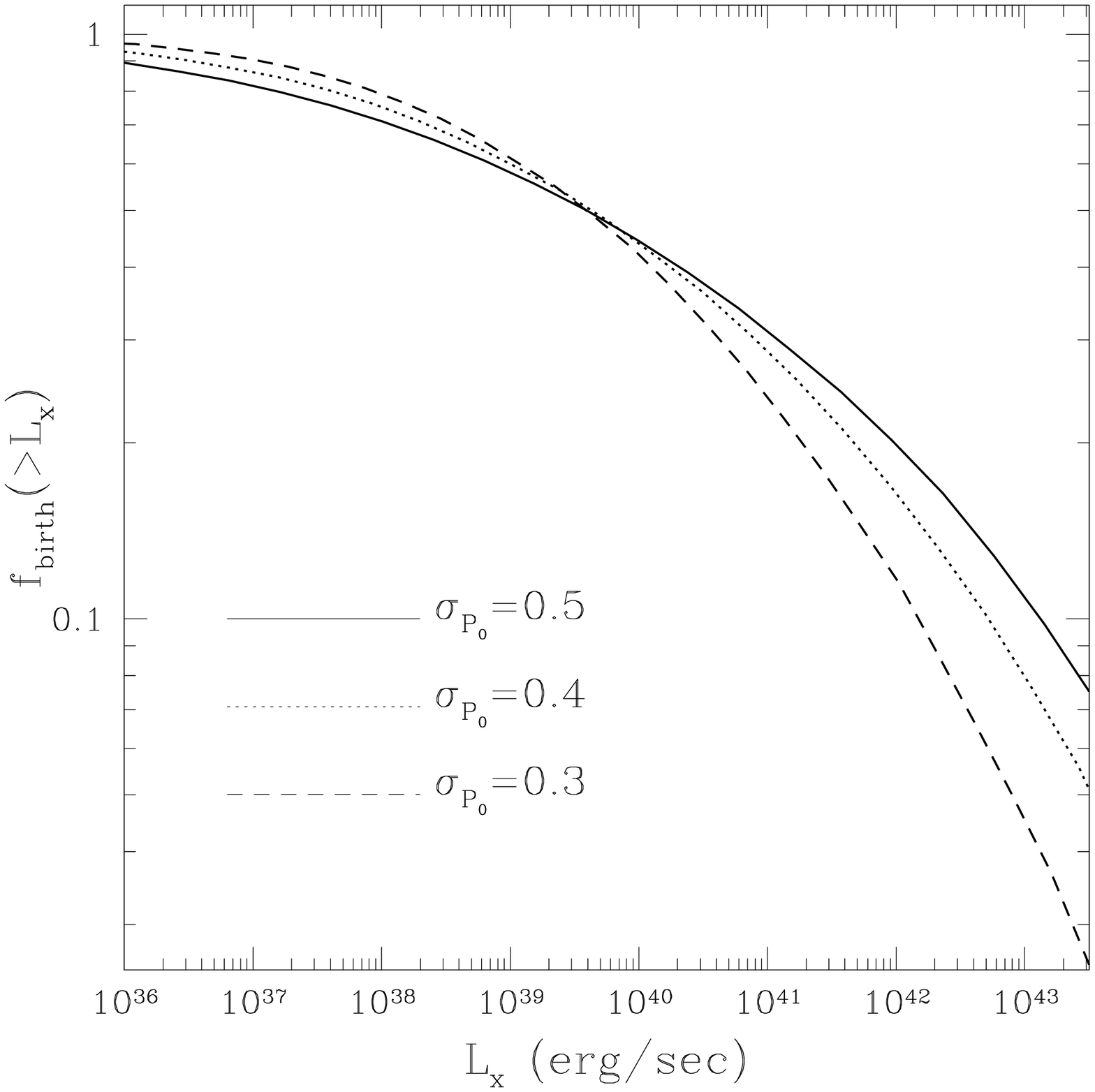}
\caption{Probability that a rotation-powered pulsar has an X-ray
luminosity at birth larger than $L_x$ for the pulsar birth parameters
of the Arzoumanian et al. (2002) distribution, and for various values
of the spread $\sigma_{P_0}$ of the initial period distribution.  
A fraction of the rotational energy is assumed to 
be converted in X-rays (in the 2-10 keV band) according to the
relation found by Possenti et al. (2002).}
\end{figure}

\begin{figure}[t]
\plotone{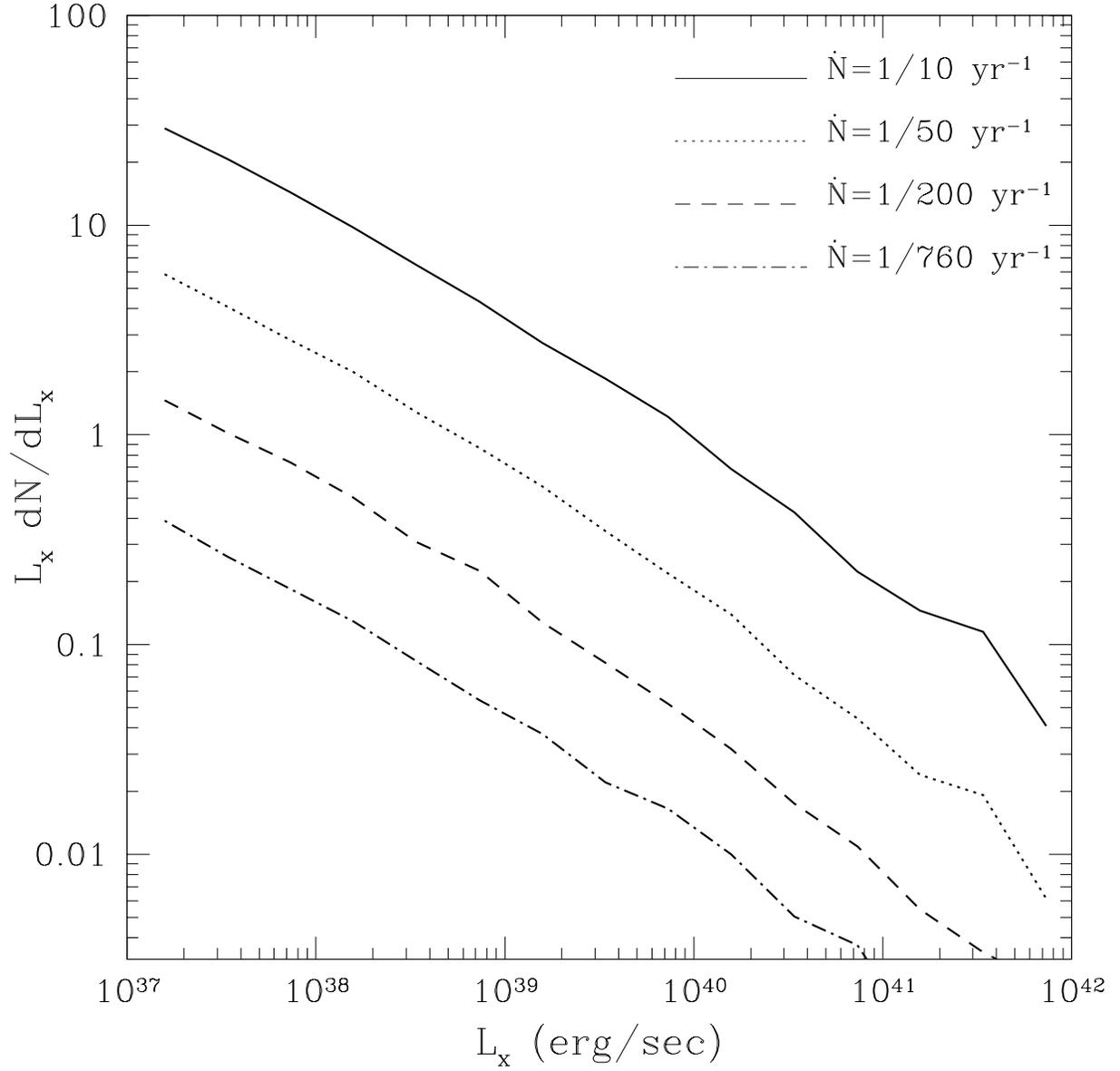}
\caption{X-ray luminosity function (2-10 keV range) of the young pulsar population
for different values of the pulsar birth rate.}
\end{figure}

\begin{figure}[t]
\plotone{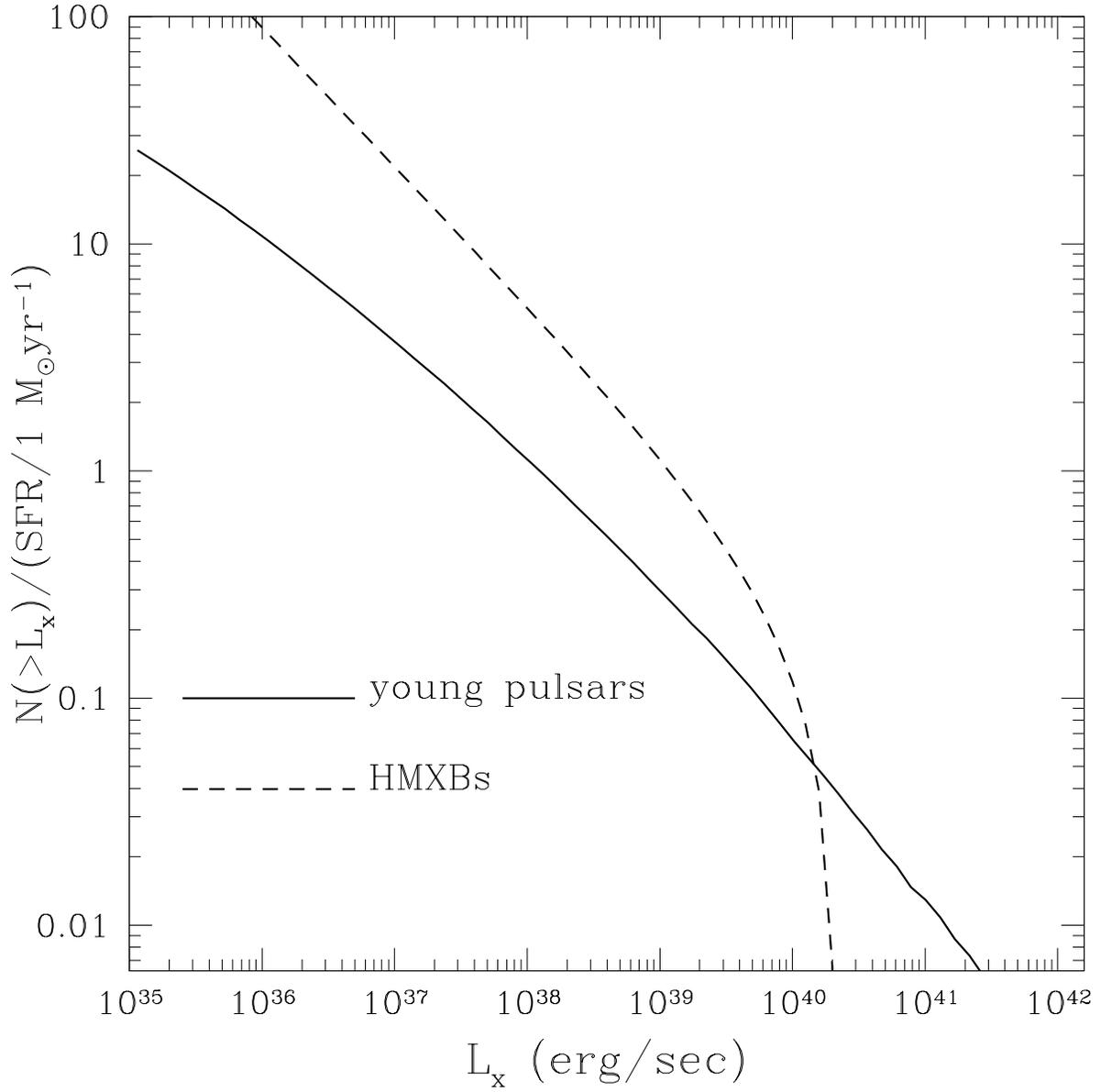}
\caption{Comparison between the pulsar and HMXB number counts,
normalized to a SFR of 1$M_\odot$/yr. }
\end{figure}

\begin{figure}[t]
\plotone{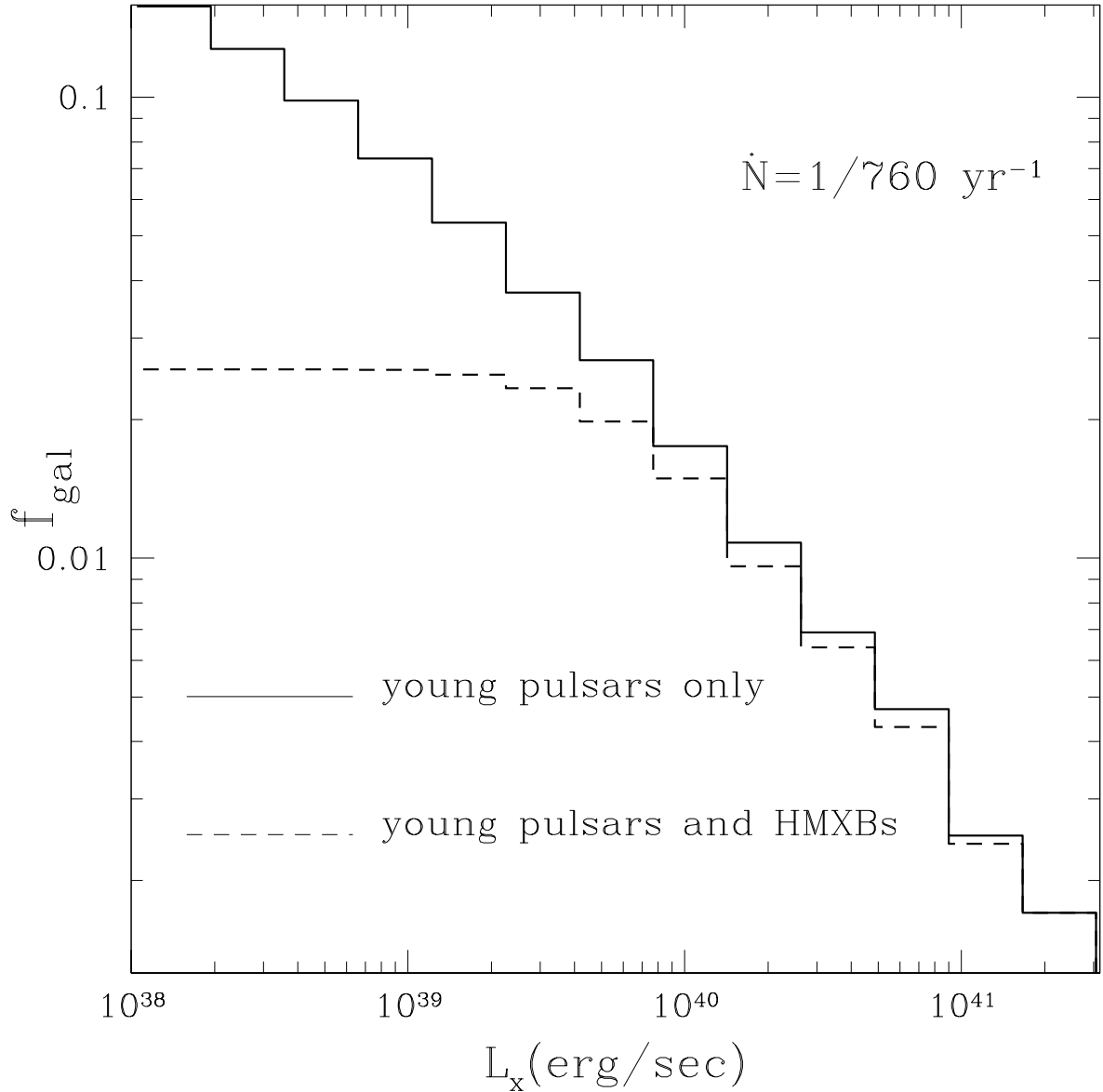}
\caption{Fraction of galaxies whose X-ray luminosity (in the 2-10 keV range) 
is dominated by a single young pulsar source with luminosity larger
than $L_x$. The condition is that the X-ray luminosity
of that single pulsar has to be $\ga 90\% L_{\rm tot}$.}
\end{figure}

\end{document}